\documentclass[journal]{IEEEtran}
\usepackage{times,amsfonts,bbm,dsfont,amsmath,color,amssymb,graphicx,epsfig,multirow,float,algorithm,algorithmic,bm,cite,setspace}
\usepackage[normalem]{ulem}
\usepackage{makecell}
\usepackage{stfloats}
\usepackage{arydshln}
\usepackage{psfrag}
\usepackage{epstopdf}
\usepackage{stfloats}
\usepackage{color,cases}
\usepackage{subfigure}
\usepackage[english]{babel}

\begin{document}

\title{Deep Reinforcement Learning based Modulation and Coding Scheme Selection in Cognitive Heterogeneous Networks}
\author{\authorblockN{Lin Zhang, Junjie Tan, Ying-Chang Liang, \emph{Fellow, IEEE}, Gang Feng, and Dusit Niyato, \emph{Fellow, IEEE} }


\thanks{L. Zhang, J. Tan, and G. Feng are with the Key Laboratory on Communications, and also with the Center for Intelligent Networking and Communications (CINC), University of Electronic Science and Technology of China (UESTC), Chengdu, China (emails: linzhang1913@gmail.com, tan@kuspot.com, and fenggang@uestc.edu.cn).}

\thanks{Y.-C. Liang is with the Center for Intelligent Networking and Communications (CINC), University of Electronic Science and Technology of China (UESTC), Chengdu, China (email: liangyc@ieee.org).}

\thanks{D. Niyato is with the School of Computer Science and Engineering, Nanyang Technological University, Singapore (email: dniyato@ntu.edu.sg).}}

  \maketitle



\begin{abstract}

We consider a cognitive heterogeneous network (HetNet), in which multiple pairs of secondary users adopt sensing-based approaches to coexist with a pair of primary users on a certain spectrum band. Due to imperfect spectrum sensing, secondary transmitters (STs) may cause interference to the primary receiver (PR) and make it difficult for the PR to select a proper modulation and/or coding scheme (MCS). To deal with this issue, we exploit deep reinforcement learning (DRL) and propose an intelligent MCS selection algorithm for the primary transmission. To reduce the system overhead caused by MCS switching¡¯s, we further introduce a switching cost factor in the proposed algorithm. Simulation results show that the primary transmission rate of the proposed algorithm without the switching cost factor is $90\%\sim 100\%$ of the optimal MCS selection scheme, which assumes that the interference from the STs is perfectly known at the PR as prior information, and is $30\%\sim 100\%$ higher than those of the benchmark algorithms. Meanwhile, the proposed algorithm with the switching cost factor can achieve a better balance between the primary transmission rate and system overheads than both the optimal algorithm and benchmark algorithms.

\end{abstract}


%
\section{Introduction}
Fueled by the exponential growth of smart phones and tablets, recent years have witnessed an explosive increase of data traffics in wireless networks \cite{IoT}, \cite{Lin_IoT_2018}. It is envisioned that wireless data traffics will continue increasing in the next few years. To accommodate these traffics, it is urgent to improve the network capacity. Two typical approaches to improve the network capacity include enhancing the wireless link efficiency and optimizing the network architecture. Nevertheless, the wireless link efficiency is approaching the fundamental limit with the development of the multiple-input-multiple-output and orthogonal frequency division multiplexing technologies. As such, a \emph{heterogeneous network} (HetNet) is emerging as a promising network architecture to improve the network capacity \cite{HetNet_survey}-\cite{HetNet_survey_1} .

Different from a conventional cellular network, the HetNet typically consists of a macro base station (BS), multiple small cell BSs, and numbers of users \cite{HetNet_survey_1}. The macro BS is deployed to provide a wide coverage for users with low data-rate requirements and the small cell BSs are to extend the coverage of the macro BS as well as to support high data-rates for the users in a relatively small area. In the deployment of the HetNet, one major challenge is the coexistence among multiple wireless links of different users. On the one hand, if dedicated spectrum bands are assigned to different wireless links to avoid interference, large amount of spectrum resource is required to satisfy massive transmission demands in the network. On the other hand, if all the wireless links share the same spectrum band, different wireless links may cause severe interference to each other. To deal with this issue, the cognitive radio technology has been introduced to the HetNet, namely, cognitive HetNet \cite{HetNet_CR_1}-\cite{HetNet_CR_4}. In particular, the cognitive HetNet consists of two types of users, i.e., primary users with high priorities to the spectrum bands and secondary users with low priorities to the spectrum bands.

%
%

To protect primary transmissions, the \emph{secondary transmitter} (ST) usually adopts a sensing-based approach to determine whether to access a target spectrum band or not. In particular, the ST first measures the energy of the signal on the target spectrum band. If the measured energy exceeds a certain threshold, the target spectrum band is declared to be occupied by primary users and the ST keeps silent. Otherwise, the target spectrum band is idle and the ST can access it directly. However, the complicated environment in a cognitive HetNet may lead to imperfect spectrum sensing.

In fact, the imperfect spectrum sensing issue commonly exists in a cognitive network. To reduce the impact of imperfect spectrum sensing on the primary transmission performance, the authors of \cite{YC_2008} suggested guaranteeing a high detection probability, e.g., $90\%$, for a relatively low strength of the received primary signal at the ST, e.g., the signal-to-noise ratio (SNR) of the received primary signal at the ST is as low as $-15$ dB. This method can is able to reduce effectively the impact of imperfect spectrum sensing on the primary transmission performance and thus is widely adopted in cognitive networks \cite{Lin_SS_2017}.

\subsection{Motivations}

It is clear that the effectiveness of the method in \cite{YC_2008} diminishes for the scenario in which the PT is transparent to the ST, i.e., the strength of the received primary signals at the ST is extremely low (much lower than $-15$ dB), and the channel from the ST to the PR is non-ignorable. In this scenario, the ST may cause severe interference to the PR after imperfect spectrum sensing and degrade the primary transmission performance. This scenario is particularly relevant to the uplink transmission of a cognitive HetNet, in which a PT transmits uplink data to the Macro BS (PR) on a certain spectrum band, and multiple pairs of secondary users adopt a sensing-based approach to coexist with the primary users on the same spectrum band. Since the antenna height of the user terminal is relatively low while that of the macro BS is high, the wireless links between the PT and STs may be heavily blocked by buildings while the line of sight propagations exist between the PR and the STs. To avoid severe interference from the STs to the primary transmission, existing literature suggested each ST adopt a conservative transmit power. However, the question still lies in whether the primary transmission performance and the network capacity can be further enhanced.

We notice that the starting time of the secondary transmission is later than that of the primary transmission according to the sensing-based protocol. As such, the interference from STs is unknown at the PT at the starting time of the primary transmission, and the PT cannot adapt its transmission with the interference information. In fact, the interference from STs typically follows a certain pattern, and it is possible for the PT to learn the interference pattern by analyzing the historical interference information and infer the interference in the future frames. In this paper, we adopt the deep reinforcement learning (DRL) for the PR to learn the interference pattern from STs and infer the interference in the future frames \cite{DRL_Survey}. With the inferred interference, the PT can adapt its transmission to enhance the primary transmission rate as well as the network capacity. In the following, we first provide related work on the applications of both RL and DRL in wireless communications and then elaborate the contributions of the paper.

\subsection{Related work}

Recently, RL is widely applied in wireless communication networks, especially in decision-making scenarios \cite{RL_0}-\cite{RL_8}. Specifically, \cite{RL_0} proposed two RL-based user handoff algorithms in a Millimeter wave HetNet. \cite{RL_1} developed an efficient RL-based radio access technology selection algorithm in a HetNet. \cite{RL_2} studied the energy-efficiency in a HetNet, and proposed a RL-based user scheduling and resource allocation algorithm. \cite{RL_3} and \cite{RL_4} investigated the spectrum sharing problem in cognitive radio networks and developed RL-based spectrum access algorithms for cognitive users. \cite{RL_5} and \cite{RL_6} focused on the self-organization network and adopted RL to deal with the request coordination problem and the user scheduling problem, respectively. In addition, \cite{RL_7} applied RL in the physical layer security and proposed an RL-based spoofing detection scheme. \cite{RL_8} formulated the wireless caching as an optimal decision-making problem and developed an RL-based caching scheme to reduce the energy consumption.

It is proved that RL works well in decision-making scenarios when the size of the state-action space in the wireless system is relatively small. However, the effectiveness of RL diminishes as the size of the state-action space becomes large. Then, DRL emerges as a good alternative to solve the decision-making problem in wireless systems with a large size of state-action space  \cite{DRL_1}-\cite{DRL_7}. In particular, \cite{DRL_1} developed a DRL-based user scheduling algorithm to enhance the sum-rate in a wireless caching network. \cite{DRL_2} proposed a DRL-based channel selection algorithm to improve the transmission performance in a multi-channel wireless network. \cite{DRL_3} adopted DRL to learn the jamming pattern in a dynamic and intelligent jamming environment and proposed an efficient algorithm to obtain the optimal anti-jamming strategy. \cite{DRL_4} adopted DRL to learn the power adaption strategy of the primary user in a cognitive network, such that the secondary user is able to adaptively control its power and satisfy the required quality of services of both primary and secondary users. \cite{DRL_5} studied the handover problem in a multi-user multi-BS wireless network and proposed a DRL-based handover algorithm to reduce the handover rate of each user under a minimum sum-throughput constraint. In addition, \cite{DRL_6} proposed a distributed DRL-based multiple access algorithm to improve the uplink sum-throughput in a multi-user wireless network. \cite{DRL_7} applied DRL to the power allocation problem in an interference channel and proposed a DRL-based algorithm to enhance the sum-rate.

\subsection{Contributions of the paper}

In this paper, we consider a cognitive HetNet, in which a mobile user (PT) transmits uplink data to the macro BS (PR) on a certain spectrum band and multiple STs adopt a sensing-based approach to access the same spectrum band. In particular, the PT is transparent to the STs, and the channel from each ST to the PR is non-ignorable. As a result, each ST may access the spectrum band with imperfect spectrum sensing and cause interference to the PR. Since the interference is unknown at the PT due to time causality, it is difficult for the PR to select a proper modulation and/or coding scheme (MCS) to improve the primary transmission performance. Note that MCS refers to modulation and coding scheme in a coded system, and is reduced to modulation scheme in an uncoded system. For consistency, we use MCS to represent modulation and coding scheme in a coded system, and represent modulation scheme in an uncoded system. We summarize the major contributions of the paper as follows:
\begin{enumerate}
\item We propose an intelligent DRL-based MCS selection algorithm for the PR. Specifically, we enable the DRL agent at the PR to learn the interference pattern from the STs. With the learnt interference pattern, the PR can infer the interference from the STs in the future frames and adaptively select a proper MCS to enhance the primary transmission rate.

\item We take the system overhead caused by MCS switchings into consideration and introduce a switching cost factor in the proposed algorithm. By adjusting the value of the switching cost factor, we can achieve different balances between the primary transmission rate and system overheads.

\item Simulation results show that the transmission rate of proposed algorithm without the switching cost factor is $90\%\sim 100\%$ to that of the optimal MCS selection scheme and is $30\%\sim 100\%$ higher than those of benchmark algorithms. Meanwhile, the proposed algorithm with the switching cost factor can achieve a better balance between the transmission rate and system overheads than those of both the optimal algorithm and benchmark algorithms.

\end{enumerate}

\subsection{Organization of the Paper}
The rest of the paper is organized as follows: Section II describes
the system model. Section III analyzes the optimal MCS selection policy. In Section IV, we elaborate the proposed intelligent DRL-based MCS selection algorithm. Section V provides simulation results to demonstrate the performance of the proposed algorithm. Finally, Section VI concludes the paper.


\section{System Model}
          \begin{figure}
            \centering
            \includegraphics[scale=1]{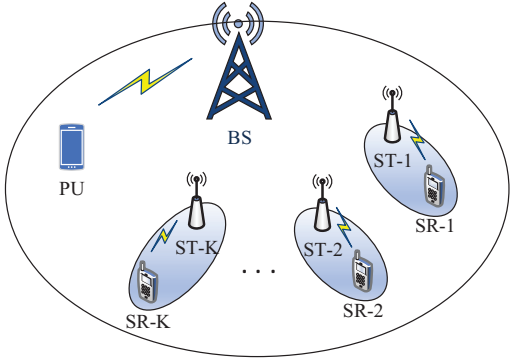}
            \caption{Considered cognitive HetNet, which consists of a PU, a macro BS, and $K$ pairs of secondary users.}
             \label{System_model}
        \end{figure}

We consider a cognitive HetNet as shown in Fig. \ref{System_model}, in which $K$ pairs of secondary users coexist with a pair of primary users in an overlay spectrum access mode. In particular, a \emph{primary user} (PU) is transmitting uplink data to a macro BS (PR) on a certain spectrum band. To protect primary transmissions, each ST (namely, ST-$k$, $k\in \{1, 2, \ldots, K\}$) adopts a sensing-based approach to determine whether to access the spectrum band and transmit data to the associated SR (namely, SR-$k$, $k\in \{1, 2, \cdots, K\}$). In the following, we provide the channel model, the coexistence model, and the signal model of the PU transmission in the considered network.

\subsection{Channel model}
Each channel in the considered network is composed of a large-scale path-loss and a small-scale block Rayleigh fading \cite{Lin_SPC_2018}. If we denote $\bar g_{\text{p}}$ as the large-scale path-loss component and denote $h_{\text{p}}$ as the small-scale block Rayleigh fading component between the PU and the BS, the corresponding channel gain is $g_{\text{p}}=\bar g_{\text{p}}|h_{\text{p}}|^2$. Similarly, if we denote $\bar g_{k}$ as the large-scale path-loss component and denote $h_{\text{p}}$ as the small-scale block Rayleigh fading component between ST-$k$ and the BS, the corresponding channel gain is $g_{k}=\bar g_{k}|h_{k}|^2$.

In particular, the large-scale path-loss component remains constant for a given distance between the corresponding transmitter and receiver, the small-scale block Rayleigh fading component remains constant in each transmission frame and varies in different transmission frames. According to \cite{Jake_model}, we adopt the Jake's model to represent the relationship between the small-scale Rayleigh fadings in two successive frames, i.e.,
\begin{align}
h(t)=\rho h(t-1)+\delta,
\end{align}
where $\rho$ is the correlation coefficient of two successive small-scale Rayleigh fading realizations, $\delta$ is a random variable with a distribution $\delta \sim \mathcal{CN}(0, 1-\rho^2)$, and $h(0)$ is a random variable with a distribution $h(0) \sim \mathcal{CN}(0, 1)$.



\subsection{Coexistence model}

          \begin{figure}
            \centering
            \includegraphics[scale=0.6]{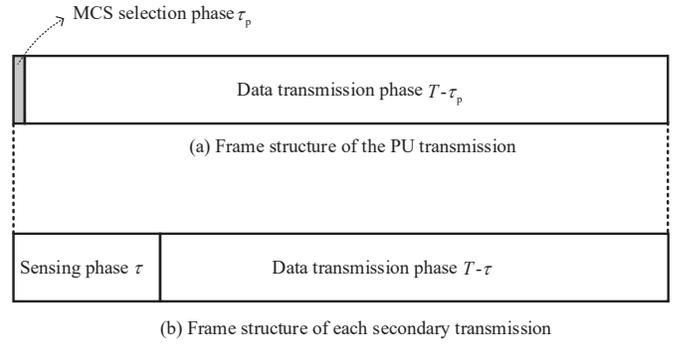}
            \caption{Frame structures: (a) the frame structure of the PU transmission; (b) the frame structure of each secondary transmission.}
             \label{Frame_structure}
        \end{figure}
As shown in Fig. \ref{Frame_structure}-(a), the duration of each PU transmission frame is $T$ and each frame is divided into two successive phases, i.e., MCS selection phase and data transmission phase. If we denote $\tau_{\text{p}}$ as the duration of the MCS selection phase, the duration of the data transmission phase is $T-\tau_{\text{p}}$. In practical situations, $\tau_{\text{p}}$ is small compared with $T$ and thus can be neglected in the system design. In the MCS selection phase, the PU first transmits training signals to the BS. By receiving the training signals, the BS estimates the channel from the PU to the BS and meanwhile measures the \emph{signal to noise ratio} (SNR). According to the measured SNR, the BS selects a proper MCS scheme and feeds it back to the PU. In the data transmission phase, the PU adopts the MCS scheme selected by the BS for the uplink data transmission and the BS uses the estimated channel information to recover the required data.

As aforementioned, secondary users need to protect primary transmissions when coexisting with primary users on the same channel. For this purpose, each ST adopts a sensing-based approach to determine whether to access the spectrum band or not \cite{HetNet_CR_Lin}. Specifically, the frame structure of the secondary transmission is synchronous with the primary transmission frame as shown in Fig. \ref{Frame_structure}-(b). In particular, the secondary transmission frame consists of two successive phases: spectrum sensing phase and data transmission phase. If we denote $\tau$ as the duration of the spectrum sensing phase, the duration of the data transmission phase is $T-\tau$. In the spectrum sensing phase, ST-$k$ senses the channel and determines whether to access the channel or not. If the channel is idle, ST-$k$ transmits data to SR-$k$ for the rest of the frame. Otherwise, ST-$k$ keeps silent. In fact, ST-$k$ also needs to select an MCS for the transmission of ST-$k$ once ST-$k$ determines to access the channel. To focus on the primary transmission design, we will omit the discussion of secondary transmissions in the paper.


Due to imperfect spectrum sensing, ST-$k$ may access the channel even when the channel is occupied by the PU transmission. Then, the PU transmission is interference-free for the former duration $\tau$ of a frame and may be interfered with by ST-$k$ for the later duration $T-\tau$ of the frame. In the rest of the paper, we denote $\alpha_k$ as the miss-detection/interference probability that ST-$k$ accesses the channel and causes interference to the primary transmission.

\subsection{Signal model of the PU transmission}




Note that, the PU transmission is not interfered with by STs in the former duration $\tau$ of each frame. If we denote $p_{\text{p}}$ as the fixed transmit power of the PU\footnote{According to \cite{MCS_SC}, power adaptation in addition to MCS adaptation gives negligible additional gains when the number of MCS levels is high. Thus, we consider a fixed transmit power of the PU and focus on the MCS adaption.}, the received SNR at the BS is
\begin{align}
    \gamma_0=\frac{p_{\text{p}}g_{\text{p}}}{\sigma^2}.
    \label{SNR}
\end{align}

In the later duration $T-\tau$ of each frame, the PU transmission may be interfered by active STs. If we denote $S_{\text{a}}$ as the set of active STs and denote $p_k$ as the fixed transmit power of ST-$k$, the received \emph{signal to interference and noise ratio} (SINR) at the BS is
\begin{align}
    \gamma_1=\frac{p_{\text{p}}g_{\text{p}}}{\sum_{k\in S_{\text{a}}}p_kg_k+\sigma^2}.
    \label{SINR}
\end{align}

Suppose that the PU adopts a bit-interleaver to cope with the burst interference from secondary transmissions. Then, the average SINR of each bit at the BS is
\begin{align}
    \bar \gamma=\frac{\tau-\tau_{\text{p}}}{T-\tau_{\text{p}}}\gamma_0+\frac{T-\tau}{T-\tau_{\text{p}}}\gamma_1.
    \label{Average_SINR}
\end{align}

\section{Optimal MCS selection policy}

In this section, we present the optimal MCS selection policy at the BS. We first provide the basic principle of the MCS selection. Then, we formulate the MCS selection as an optimization problem and elaborate the optimal policy.

\subsection{Basic principle}
Basically, for a given average SINR, there exists a tradeoff between the MCS and the transmission rate of the packet in a frame. Specifically, a low-order MCS leads to a low \emph{symbol error rate} (SER) as well as a low \emph{packet error rate} (PER), which improves the transmission reliability and consequently the transmission rate. Meanwhile, a low-order MCS corresponds to a low transmission rate. Thus, it is necessary to select the optimal MCS to maximize the transmission rate. Note that, the MCS selection can be performed at either the PU or the BS \cite{MCS_SC}. In this paper, the MCS selection is performed at the BS for two main reasons. Firstly, the computing capability of the BS is typically stronger than the PU. Secondly, the BS can directly measure and analyze the SINR, which contains the interference information from STs.

\subsection{The optimal modulation policy}

Suppose that $M$ levels of MCSs (namely, MSC$_m$ ($m\in \{1, 2,\cdots, M\}$)) are available for the uplink transmission from the PU to the BS. In particular, we denote $f_m(\bar\gamma)$ as the \emph{symbol error rate} (SER) of MSC$_m$, and denote $r_m$ (bits/symbol) as the average transmission efficiency of MSC$_m$.

Denote $N$ as the number of transmitted symbols in each primary packet/frame. A packet error happens if any transmitted symbol is not correctly decoded by the BS. Accordingly, the packet error rate of the primary transmission can be approximated as \cite{PER}
\begin{align}
    \rho_m(\bar \gamma)\approx 1-\left(1-f_m(\bar \gamma)\right)^{N}.
    \label{PER_i}
\end{align}
Then, the transmission rate (bits/frame) can be written as
\begin{align}
    R_m(\bar \gamma)=r_m[1-\rho_m(\bar \gamma)]N.
    \label{PER_i}
\end{align}

The optimal MCS selection policy aims to select the optimal MCS in the MCS selection phase to maximize the transmission rate in each frame, i.e.,
\begin{align}
    m^*=\arg \max_{m \in \{1, 2, \cdots, M\}} R_m (\bar \gamma).
    \label{Optimization_problem}
\end{align}

It is clear that the optimization problem (\ref{Optimization_problem}) can be solved by two steps: In the first step, the BS calculates $R_m(\bar \gamma)$ for each MCS$_m$, $m\in \{1, 2, \cdots, M\}$; in the second step, the BS selects the optimal MCS index $m^*$ subject to the maximum $R_{m^*}(\bar \gamma)$. Note that $\bar \gamma$ is needed to complete the two steps. According to (\ref{Average_SINR}), $\bar \gamma$ is determined by $\gamma_0$ and $\gamma_1$. In particular, $\gamma_0$ can be directly obtained by the BS in the MCS selection phase of each frame. $\gamma_1$ is related to the interference from secondary transmissions in the data transmission of each frame and is unknown at the BS in the MCS selection phase of each frame due to the time casualty. In other words, the BS cannot obtain $\bar \gamma$ in the MCS selection phase of each frame. Thus, it is impractical for the BS to select the optimal MCS by solving the optimization problem (\ref{Optimization_problem}).

A straightforward MCS selection policy at the BS is ignoring the interference from STs and selecting an MCS based on the SNR $\gamma_0$ of the received training signal in the MCS selection phase. In particular, by replacing $\bar \gamma$ in the optimization problem (\ref{Optimization_problem}) with $\gamma_0$, we can obtain a straightforward MCS selection policy. Nevertheless, this MCS selection policy is suboptimal in terms of the transmission rate since the interference from secondary users is not considered.

\section{Intelligent DRL-based MCS selection algorithm}

In this section, we propose an intelligent DRL-based MSC selection algorithm for the BS to select the optimal MCS and maximize the transmission rate from the PU to the BS. Next, we provide the basic principle followed by the algorithm development.

\subsection{Basic principle}


As aforementioned, the optimal MCS selection is calculated by $\bar \gamma$, which depends on $\gamma_1$. Since $\gamma_1$ is determined by the interference from secondary transmissions in the data transmission phase of each frame, it is impractical for the BS to calculate the optimal MCS selection policy in the MCS selection phase of each frame due to the time casualty. In fact, the interference from secondary transmissions usually follows a certain pattern. Specifically, the interference from secondary transmissions is mainly determined by two factors: the transmit power of each ST and the channel gain from each ST to the BS. Note that neither of the two factors is known to the BS, resulting in that the interference pattern is hidden to the BS. Nevertheless, the interference from secondary transmissions can be measured by the BS at the end of each data transmission phase. Therefore, it is possible for the BS to learn the hidden interference pattern by collecting and analyzing the historical interference from secondary transmissions. With the learnt interference pattern, the BS is able to infer the interference from STs in the data transmission phase and select a proper MCS to maximize the transmission rate in each frame.

In fact, the optimal MCS selection is an optimal decision-making problem. From \cite{DRL_nature}, DRL is an effective tool to learn a hidden pattern in a decision-making problem and gradually achieves the optimal policy via trail-and-error. Therefore, we can adopt DRL to learn the interference pattern of STs and design the optimal MCS selection policy to maximize the transmission rate. Since DRL originates from RL, we will first provide both the general RL framework and the general DRL framework, and then elaborate the proposed intelligent DRL-based MCS selection algorithm.

\subsection{General RL framework}
Basically, there are six fundamental elements in a general RL framework, i.e., action space $\textbf{\emph{A}}$, state space $\textbf{\emph{S}}$, immediate reward $r(s, a)$, ($s\in \textbf{\emph{S}}, a \in \textbf{\emph{A}}$), transition probability space $\textbf{\emph{P}}$, Q-function $\textbf{\emph{Q}}(s, a)$, ($s\in \textbf{\emph{S}}, a \in \textbf{\emph{A}}$), and policy $\pi$. Specifically,

\begin{enumerate}

\item Action space $\textbf{\emph{A}}$: the action space is a set of all the actions $a$ that are available for the RL agent to select;

\item State space $\textbf{\emph{S}}$: the state space is a set of all the environmental states $s$ that can be observed by the RL agent;

\item Immediate reward $r(s, a)$: the immediate reward $r(s, a)$ is the reward by executing the action $a$ at an environmental state $s$;

\item Transition probability space $\textbf{\emph{P}}$: the transition probability space is a set of transition probabilities $p_{ss'}(a) \in \textbf{\emph{P}}$ from an environmental state $s$ to another environmental state $s'$ after the DRL agent takes an action $a$ at the environmental state $s$;

\item Q-function $\textbf{\emph{Q}}(s, a)$: the Q-function is the expected cumulative discounted reward in the future (namely, long-term reward) by executing the action $a$ at an environmental state $s$;

\item Policy $\pi(s)\in \textbf{\emph{A}}$: the policy is a mapping from the environmental states observed by the RL agent to the actions that will be selected by the RL agent.
\end{enumerate}

Typically, the transition probability of the environmental state is impacted by both the environment itself and the action of the RL agent, and thus is modelled as a \emph{markov decision process} (MDP). By introducing immediate rewards of state-action pairs in the MDP, the RL agent is stimulated to adopt the action policy that maximizes the long-term reward. Note that the maximization of the long-term reward is not equivalent to the maximization of the immediate reward. For instance, a state-action pair producing a high immediate reward may have a low long-term reward because the state-action pair may be followed by other states that yield low rewards. In other words, the long-term reward of a state-action pair is related to not only its immediate reward but also the future rewards. Thus, to obtain the optimal action policy that maximizes the long-term reward, the RL agent needs to determine the long-term reward of each state-action pair by analyzing a long sequence of state-action-reward pairs that follows each state-action pair.

Since the long-term reward is related to both its immediate reward and the future rewards, we can express the long-term reward in a recursive form as follows:
\begin{align}
\textbf{\emph{Q}}(s, a) = r(s, a) + \eta \sum\limits_{s' \in \textbf{\emph{S}}} \sum\limits_{a' \in \textbf{\emph{A}}} {{p_{ss'}}(a)} \textbf{\emph{Q}}(s', a'),
\label{DRL_1}
\end{align}
where $\eta \in [0, 1]$ is the discount factor representing the discounted impact of the future reward, and $(s', a')$ is the next state-action pair after the RL agent executes the action $a$ at the environmental state $s$. The RL agent aims to find the optimal policy $\pi^*(s)$ to maximize the long-term reward $\textbf{\emph{Q}}(s, a)$ for each environmental state $s$. If we denote $\textbf{\emph{Q}}^*(s, a)$ as the highest long-term reward for the state-action pair $(s, a)$, we can rewrite (\ref{DRL_1}) as
\begin{align}
\textbf{\emph{Q}}^*(s,a) = r(s,a) + \eta \sum\limits_{s' \in \textbf{\emph{S}}} {{p_{ss'}}(a)\mathop {\max }\limits_{a' \in \textbf{\emph{A}}} } \ \textbf{\emph{Q}}^*(s',a').
\label{DRL_2}
\end{align}
Then, the optimal action policy is
\begin{align}
{\pi ^*}(s) = \mathop {\arg \max }\limits_{a \in \textbf{\emph{A}}} \left[ {{\textbf{\emph{Q}}^*}(s,a)} \right], \ \forall \ s\in \textbf{\emph{S}}.
\end{align}

However, it is challenging to directly obtain the optimal $\textbf{\emph{Q}}(s, a)$ from (\ref{DRL_2}) since the transition probability ${p_{ss'}}(a)$ is typically unknown to the RL agent. To deal with the issue, RL agent adopts a Q-learning algorithm. In particular, the Q-learning algorithm constructs a $|\textbf{\emph{S}}| \times |\textbf{\emph{A}}|$ Q-table with the Q-function $\textbf{\emph{Q}}(s, a)$ as elements, which are randomly initialized. Then, the RL agent adopts an $\epsilon$-greedy algorithm to choose an action for each environmental state and updates each element $\textbf{\emph{Q}}(s, a)$ in the Q-table as follows:
\begin{align}
Q(s,a) \!\gets \!(1\!-\!\alpha )Q(s,a) \!+\!\alpha \!\left[\! {r(s,a) \!+\! \eta \mathop {\max }\limits_{a' \in \textbf{\emph{A}}} Q(s',a')} \!\right],
\label{Q_update}
\end{align}
where $\alpha$ is the learning rate.

The basic idea of the $\epsilon$-greedy algorithm is as follows. In general, the RL agent prefers choosing the best action that produces the highest long-term reward for each environmental state. If the RL agent has experienced the best action for a certain environmental state, it can exploit the experience to improve the long-term reward. Nevertheless, it is likely that the RL agent has not experienced the best action for the environmental state. Thus, the RL agent needs to explore the best action. To balance the exploitation of experiences and the exploration of best actions, the RL agent adopts an $\epsilon$-greedy algorithm to choose an action for each environmental state. Specifically, for a given state $s$, the RL agent executes the action $a=\arg \max_{a\in  \textbf{\emph{A}}}\textbf{\emph{Q}}(s,a)$ with the probability $1-\epsilon$, and the RL agent randomly executes an action in the action space $\textbf{\emph{A}}$ with the probability $\epsilon$. It is worth pointing out that, $\epsilon$ is a trade-off factor between the exploitation and the exploration. By optimizing $\epsilon$, RL agent can achieve the optimal policy with the fastest speed. In practical situations, $\epsilon$ is optimized through simulations.

According to the existing literature, the performance of the Q-learning algorithm differs a lot for different sizes of the state-action space. When the state-action space is small, the RL agent can experience all the state-action pairs in the state-action space rapidly and achieve the the optimal action policy. When the state-action space is large, the performance of the Q-learning algorithm diminishes since many state-action pairs may not be experienced by the RL agent and the storage size of the Q-table is unacceptably large.

\subsection{General DRL framework}


To overcome the drawback of the Q-learning algorithm in a system with a large state-action space, DRL adopts a \emph{deep Q-learning network} (DQN) $\textbf{\emph{Q}}(s,a; \theta)$, where $\theta$ is the weights of the DQN, to approximate the Q-function $\textbf{\emph{Q}}(s,a)$. In this way, by inputting the environmental state $s$ into the DQN $\textbf{\emph{Q}}(s,a; \theta)$, the DQN $\textbf{\emph{Q}}(s,a; \theta)$ outputs the long-term rewards of executing each action $a$ in $\textbf{\emph{A}}$ at the environmental state $s$. Accordingly, the optimization of $\textbf{\emph{Q}}(s,a)$ in the Q-learning algorithm is equivalent to the optimization of $\theta$ in the DQN $\textbf{\emph{Q}}(s,a; \theta)$. Meanwhile, the DRL agent learns the relationship among different environmental states and actions by continuously interacting with the environment, i.e., executing actions, receiving immediate rewards, and recording transitions of environmental states. By continuously analyzing the historical states, actions, and rewards, the DRL agent updates $\theta$ iteratively.

To update $\theta$, we define an experience of the DRL agent as $e=\left\langle {s,a,r,s'} \right\rangle$ and define a prediction error (loss function) for the experience $e$ as
     \begin{align}
\mathbb{L}(\theta ) = {{{[y^\text{Tar} - \textbf{\emph{Q}}(s,a;\theta)]}^2}} ,
\label{loss_function}
\end{align}
where $y^{\text{Tar}}$ is the target output of the DQN, i.e.,
\begin{align}
{y^\text{Tar}} = r + \eta \mathop {\max }\limits_{a' \in \textbf{\emph{A}}} \ \textbf{\emph{Q}}(s',a';\theta).
\label{trained_output}
\end{align}

To minimize the prediction error (loss function) in (\ref{loss_function}), the DRL agent usually adopts a gradient decent method to update $\theta$ once obtaining a new experience $e$. In particular, the update procedure of $\theta$ in the DQN is as follows:
\begin{align}
\theta \gets \theta- [y^\text{Tar} - \textbf{\emph{Q}}(s,a;\theta)]\nabla \textbf{\emph{Q}}(s,a;\theta).
\label{Loss_minimization_problem}
\end{align}


\subsection{Intelligent DRL-based MCS selection algorithm}

To begin with, we define the action space, state space, immediate reward function of the proposed intelligent DRL-based MCS selection algorithm as follows:

\subsubsection{Action space}
Note that the DRL agent at the BS aims to choose the optimal MCS for the PU's uplink transmission at the beginning of each frame. Thus, the action space of the DRL agent is designed to include all the available MSC levels, i.e.,
\begin{align}
\textbf{\emph{A}}=\{\text{MCS}_1, \text{MCS}_2, \cdots, \text{MCS}_M\}.
\label{Action_space_Proposed}
\end{align}
If we denote $a(t)$ as the selected action of the DRL agent at the beginning of frame $t$, we have $a(t) \in \textbf{\emph{A}}$.

\subsubsection{Immediate reward function}
Since the objective of the DRL agent is to choose the optimal action and maximize the transmission rate from the PU to the BS, the immediate reward of an action shall be proportional to the amount of data bits that have been successfully transmitted from the PU to the BS. Therefore, we define the immediate reward function as the number of transmitted data bits if the transmission is successful and zero otherwise, i.e.,
\begin{align}
r(t)=\left\{
\begin{aligned}
r_mN, & \quad \text{if successful,} \\
0,  & \quad \text{if failed.}
\end{aligned}
\right.
\label{Reward_function}
\end{align}

\subsubsection{State space}
The DRL agent updates the action policy by analyzing the experiences $e=\left\langle {s,a,r,s'} \right\rangle$ and chooses a proper action at the beginning of a frame based on the current state. To maximize the transmission rate, each state is supposed to provide some useful knowledge for the DRL agent to choose the optimal MCS. We notice that the optimal MCS selection is related to three types of information. Firstly, the optimal MCS in a frame is related to the channel quality from the PU to the BS in the current frame. As such, the DRL agent inclines to choose a high MCS level for a strong channel from the PU to the BS, and vice versa. According to the frame structure shown in Fig. 2, the BS is able to obtain the channel quality from the PU to the BS (SNR at the BS) at the beginning of each frame. Thus, the state is designed to include the SNR of the current frame at the BS. Secondly, the optimal MCS is related to the interference from the STs in the data transmission phase. Since the BS cannot directly obtain the interference at the beginning of a frame, it is impractical to include the interference of the frame at the state. Nevertheless, the state can include some historical data for the DRL agent to learn the interference pattern from the STs, such that the DRL agent can infer the future interference from STs. Thus, the state of a frame is also designed to include both the SNR and the SINR at the BS in the previous $\Phi$ frames. Thirdly, the optimal MCS is related to the rule that maps the optimal action from the former two types of information. Since the information of the mapping rule is contained in the historical channel quality from the PU to the BS, the historical interference from STs, the historical actions, and the historical rewards, the state at the beginning of a frame is also designed to include the action and its immediate reward in the previous $\Phi$ frames. To summarize, we define the state in frame $t$ as
\begin{align}\nonumber
s(t)=\{&a(t-\Phi), r(t-\Phi), \gamma_0(t-\Phi), \bar \gamma(t-\Phi),\ldots, \\
       &a(t-1), r(t-1), \gamma_0(t-1), \bar \gamma(t-1), \gamma_0(t)\}.
\label{state_definition}
\end{align}

We present the structure of the proposed intelligent DRL-based MCS selection algorithm in Fig. 3, which consists of the flow charts of the the cognitive HetNet in the MCS selection phase and the data transmission phase. We mainly consider four functional modules at the BS, i.e., signal transmitting/receiving module, DRL agent, local memory $\mathbb{D}$, and $\epsilon$-greedy algorithm module.

In the MCS selection phase of frame $t$, the PU first transmits pilot signals to the signal transmitting/receiving module at the BS. By receiving the pilot signals, the signal transmitting/receiving module measures the SNR $\gamma_0(t)$ and forwards it to the DRL agent. Then, the DRL agent observes a state $s(t)=\{a(t-\Phi), r(t-\Phi), \gamma_0(t-\Phi), \bar \gamma(t-\Phi),\ldots, a(t-1), r(t-1), \gamma_0(t-1), \bar \gamma(t-1), \gamma_0(t)\}$ and forms an experience $e(t)=\left\langle {s(t-1),a(t-1),r(t-1),s(t)} \right\rangle$. After that, the DRL agent stores the experience in the local memory $\mathbb{D}$, and subsequently inputs $s(t)$ to the $\epsilon$-greedy algorithm module, which outputs the selected action $a(t)$ to the signal transmitting/receiving module. To this end, the signal transmitting/receiving module feeds the selected $a(t)$ back to the PU.

In the data transmission phase of frame $t$, the PU transmits signals to the signal transmitting/receiving module at the BS in the presence of the interference from the STs. By receiving both the signals and interference, the signal transmitting/receiving module measures the average SINR $\bar \gamma(t)$ and observes the corresponding immediate reward $r(t)$. Finally, the DRL randomly chooses experience samples to train the DQN, i.e., update the weights therein.

The pseudocode of the proposed intelligent DRL-based MCS selection algorithm is shown in Algorithm 1. In particular, besides the general DRL framework, the proposed intelligent DRL-based MCS selection algorithm adopts ``experience replay'' and ``quasi-static target network'' techniques in \cite{DRL_nature} for the stabilization. For the experience replay, once obtaining a new experience, the DRL agent puts it into a local memory $\mathbb{D}$, which is capable of storing $N_E$ experiences, in a first-in-first-out fashion. Then, the DRL agent randomly samples a mini-batch of $Z$ experiences from the local memory $\mathbb{D}$ for the bath-training instead of training the DQN with a single experience. For the quasi-static target network, there exist two DQNs in the proposed algorithm, i.e., $\textbf{\emph{Q}}(s,a;\theta)$ and $\textbf{\emph{Q}}(s,a;\theta^-)$. In particular, $\textbf{\emph{Q}}(s,a;\theta)$ is called trained DNQ and $\textbf{\emph{Q}}(s,a;\theta^-)$ is called target DNQ. The target DNQ $\textbf{\emph{Q}}(s,a;\theta^-)$ is used to replace the trained DQN $\textbf{\emph{Q}}(s,a;\theta)$ in (\ref{trained_output}). The weights of the trained DQN are updated by the weights of the target DQN every $L$ frames. Accordingly, the loss function and the update procedure of $\theta$ from (\ref{loss_function}) to (\ref{Loss_minimization_problem}) can be replaced by
\begin{align}
\mathbb{L}(\theta ) = \frac{1}{2Z}\sum_{e\in E}{{{[y^\text{Tar} - \textbf{\emph{Q}}(s,a;\theta)]}^2}} ,
\label{loss_function_1}
\end{align}
where
\begin{align}
{y^\text{Tar}} = r + \eta \mathop {\max }\limits_{a' \in \textbf{\emph{A}}} \ \textbf{\emph{Q}}(s',a';\theta^-),
\label{target_output_1}
\end{align}
\begin{align}
\theta \gets \theta- \frac{1}{Z}\sum_{e\in E}[y^\text{Tar} - \textbf{\emph{Q}}(s,a;\theta)]\nabla \textbf{\emph{Q}}(s,a;\theta).
\label{Loss_minimization_problem_1}
\end{align}

It is worth pointing out that, a general convergence proof of the DRL is still an open problem (if possible) \cite{DRL_Survey}, \cite{DRL_1}-\cite{DRL_7}, \cite{DRL_nature}. Nevertheless, the convergence of the proposed DRL-based MCS selection algorithm is validated through simulation results.

\begin{figure}[!t]
\centering
\includegraphics[scale=1]{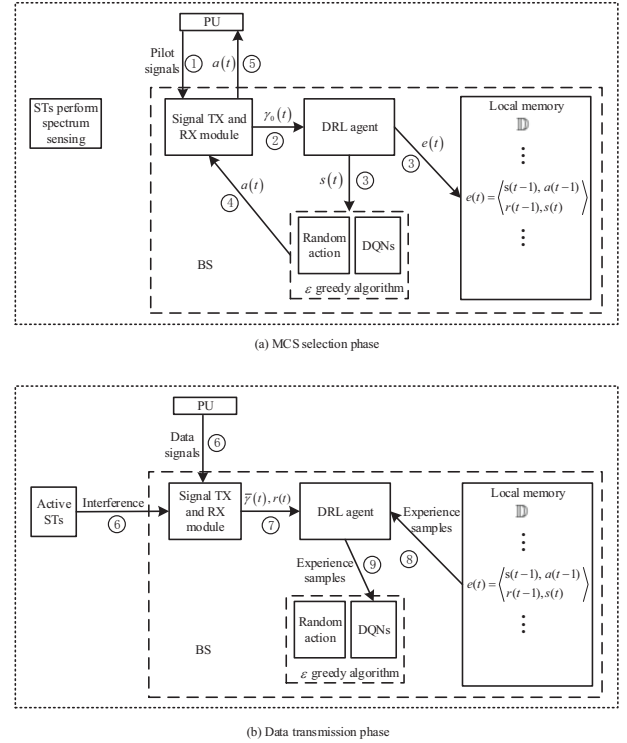}
\caption{The structure of the proposed DRL-based MCS selection algorithm.}
\label{Algorithm_diagram}
\end{figure}


 \begin{algorithm}[!thp]
\caption{Intelligent DRL-based MCS selection algorithm.}
{\small
\begin{algorithmic}[1]
\STATE Establish two DQNs (a trained DQN with weights $\theta$ and a target DQN with weights $\theta^-$).
\STATE Initialize $\theta$ randomly and enable $\theta^-=\theta$.
\STATE In frame $t$ ($t\leq Z$), the DRL agent at the BS randomly selects an action to execute and records the corresponding experience $\left\langle {s,a,r,s'} \right\rangle$ in its memory $\mathbb{D}$. Then, the DRL agent has $Z$ experiences after the first $Z$ blocks/frames.

\STATE \textbf{Repeat:}

\STATE In the block/frame $t$ ($t>Z$), the DRL agent at the BS selects an action $a(t)$ with the $\epsilon$-greedy policy: the DRL agent selects the action $a(t)=\arg \max_{a\in  \textbf{\emph{A}}}\textbf{\emph{Q}}(s(t),a; \theta)$ with the probability $1-\epsilon$, and randomly selects an action $a(t)$ in the action space with the probability $\epsilon$.

\STATE After the BS executes the selected action $a(t)$, the DRL agent obtains an immediate reward $r(s(t),a(t))$.

\STATE The DRL agent observes a new state $s(t+1)$ in frame $t+1$.

\STATE The DRL agent stores the experience $\left\langle {s(t),a(t),r(t),s(t+1)} \right\rangle$ into the local memory $\mathbb{D}$.

\STATE The DRL agent randomly samples a mini-batch with $Z$ experiences from the local memory $\mathbb{D}$.

\STATE The DRL agent updates the weights $\theta$ of the trained DQN with (\ref{Loss_minimization_problem_1}).

\STATE In every ${L}$ frames, the DRL agent updates the weights of the target DQN with ${\theta ^ -} = {\theta }$.

\end{algorithmic}}
\end{algorithm}

\subsection{Intelligent DRL-based MCS selection algorithm with switching costs}

Due to the dynamic interference from the STs as well as the dynamic channel quality between the PU and the BS, the state observed by the DRL agent may vary rapidly. As such, the selected MCS by the DRL agent may switch frequently among different MCS's to maximize the long-term reward. On the one hand, since each MCS switching requires the negotiation between the PU and the BS and system reconfiguration, frequent MCS switchings may increase both signalling overheads and system reconfiguration costs \cite{MCS_SC}. On the other hand, since the information exchange of each MCS switching needs both spectrum resource and energy consumption, frequent MCS switchings may degrade both spectral and energy efficiencies.

In fact, there may exist some MCS switchings that have little impact on the long-term reward. For instance, at the beginning of frame $t-1$, the state observed by the DRL agent is $s(t-1)$ and the DQN is $\textbf{\emph{Q}}(s,a;\theta(t-2))$. Then, the selected action is $a(t-1)=\arg \max_{a\in  \textbf{\emph{A}}}\textbf{\emph{Q}}(s(t-1),a;\theta(t-2))$ and the updated DQN is $\textbf{\emph{Q}}(s,a;\theta(t-1))$. An the beginning of frame $t$, the state observed by the DRL agent is $s(t)$ and the selected action should be $a(t)=\arg \max_{a\in  \textbf{\emph{A}}}\textbf{\emph{Q}}(s(t),a;\theta(t-1))$. If $a(t)$ is different from $a(t-1)$, an MCS switching event happens. However, $\textbf{\emph{Q}}(s(t),a(t);\theta(t-1))$ may be slightly larger than $\textbf{\emph{Q}}(s(t),a(t-1);\theta(t-1))$ and the MCS switching may have little impact on the long-term reward. Thus, the MCS switching can be avoided to reduce system overheads of MCS's switchings.

To balance the long-term reward and system overheads of MCS's switchings, we introduce a switching cost factor $c$ in the immediate reward function, i.e.,
\begin{align}
r(t)\!=\!\left\{\!\!
\begin{aligned}
r_mN, & \quad \text{if} \ a(t)=a(t-1)\ \text{and successful,} \\
r_mN-c,  & \quad \text{if} \ a(t)\neq (t-1)\ \text{and successful,}\\
0, & \quad \text{if} \ a(t)= (t-1) \ \text{and failed,} \\
-c,  & \quad \text{if} \ a(t)\neq (t-1)\ \text{and failed}.
\end{aligned}
\right.
\label{Reward_function_SC}
\end{align}

In particular, $c$ represents the overall impact of an MCS switching on the system overhead and is a relative value in terms of the transmitted data bits \cite{DRL_5}, \cite{MCS_SC}. By replacing (\ref{Reward_function}) with (\ref{Reward_function_SC}) in Algorithm 1 and adjusting the value of $c$, the DRL agent can achieve different balances between the transmission rate and system overheads of MCS's switchings.

\section{Simulation results}
In this section, we provide simulation results to evaluate the performance of the proposed intelligent DRL-based MCS selection algorithm. For comparison, we consider the optimal MCS selection algorithm, which assumes that the BS knows the average SINR $\bar \gamma$ at the beginning of each frame and solves (\ref{Optimization_problem}) to obtain the optimal solution. As aforementioned, it is impractical for the BS to know the average SINR $\bar \gamma$ at the beginning of each frame due to time causality. Thus, the performance of the optimal MCS selection algorithm is the theoretical upper bound. Meanwhile, we provide two benchmark algorithms, namely, SNR-based algorithm and upper confidence bandit (UCB) learning algorithm \cite{DRL_5} \cite{UCB}. In particular, SNR-based algorithm replaces the average SINR $\bar \gamma$ in (\ref{Optimization_problem}) with the measured SNR $\gamma_0$ and solves (\ref{Optimization_problem}) to obtain a solution. The selected MCS with the UCB learning algorithm at the beginning of frame $t$ is as follows:
\begin{align}
m^*=\arg \max_{m \in \{1, 2, \cdots, M\}} \left(\mu_m+\sqrt{\frac{2\ln t}{\Gamma_m(t-1)}}\right),
\end{align}
where $\Gamma_m(t-1)$ is the number of times that MCS$_m$ has been selected in the previous $t-1$ frames, $\mu_m$ is randomly initialized and updated by
\begin{align}
\mu_{m^*} \gets \mu_{m^*}+ \frac{1}{\Gamma_{m^*}(t)}\left(r(t)-\mu_{m^*}\right).
\end{align}

\begin{table}[h]
\center
\caption{Considered MCSs and the corresponding SERs.}
\footnotesize
\begin{tabular}{|c|c|}
\hline
MCS  & SER \cite{Digital_Com}\\
\hline
BPSK & $f_{1}(\bar \gamma)=Q\left(\sqrt{2 \bar \gamma}\right)$\\
\hline
QPSK & $f_{2}(\bar \gamma)=2\left(1-\frac{1}{\sqrt{4}}\right)Q\left(\sqrt{\frac{3 \log_2(4)\bar \gamma}{4-1}}\right)$ \\
\hline
16QAM & $f_{3}(\bar \gamma)=2\left(1-\frac{1}{\sqrt{16}}\right)Q\left(\sqrt{\frac{3 \log_2(16)\bar \gamma}{16-1}}\right)$\\
\hline
64QAM & $f_{4}(\bar \gamma)=2\left(1-\frac{1}{\sqrt{64}}\right)Q\left(\sqrt{\frac{3 \log_2(64)\bar \gamma}{64-1}}\right)$ \\
\hline
\end{tabular}
\end{table}

\subsection{Assumptions and settings in the simulation}

It is clear that secondary transmissions do not have any impact on the primary transmission when the primary transmission is inactive (i.e., the PU does not transmit data to the BS). When the primary transmission is active (i.e., the PU is transmitting data to the BS), the secondary transmissions may interfere with the PU transmission due to imperfect spectrum sensing. To investigate the effectiveness of the proposed algorithm in the presence of imperfect spectrum sensing, we assume that the primary transmission is always active. Then, the miss-detection/interference probability of ST-$k$ is also the active probability of ST-$k$.

In the simulation, we consider an uncoded system and assume that the PU supports four MCS levels as shown in Table I, although the proposed algorithm can be easily applied to coded systems. The DQN is composed of an input layer with $4\Phi+1$ ports, which correspond to $4\Phi+1$ elements in $s(t)$, two fully connected hidden layers, and an output layer with four ports, which correspond to four MCS levels in Table I. In particular, each hidden layer has 100 neurons with the Relu activation function. We apply an adaptive $\epsilon$-greedy algorithm, in which $\epsilon$ follows $\epsilon(t+1)=\max\{\epsilon_{\min}, (1-\lambda_{\epsilon})\epsilon(t)\}$ \cite{DRL_7}. An intuitive explanation of adopting a varying $\epsilon$ is as follows. At the beginning frames of the proposed algorithm, the number of the experienced state-action pairs is small and the DRL agent needs to explore more actions to improve the long-term reward. As the number of the experienced state-action pairs increases, the DRL agent does not need to perform so many explorations. We set $\epsilon(0)=0.3$, $\epsilon_{\min}=0.005$, and $\lambda_{\epsilon}=0.0001$. Besides, the batch size of experience samples in the proposed algorithm is $Z=32$, and the local memory at the DRL agent is $N_E=500$. Furthermore, we set $\gamma = 0.5$, and the RMSProp optimization algorithm with a learning rate $0.01$ is used to update $\theta$ \cite{RMSProp}. In addition, we set $\frac{\tau-\tau_{\text{p}}}{T-\tau_{\text{p}}}=0.1$ and $\frac{T-\tau}{T-\tau_{\text{p}}}=0.9$ in (\ref{Average_SINR}), $L=100$, and each frame contains $N=1000$ symbols.

\subsection{Performance comparison in quasi-static and dynamic interference scenarios}

\begin{figure}[!t]
\centering
\includegraphics[scale=0.6]{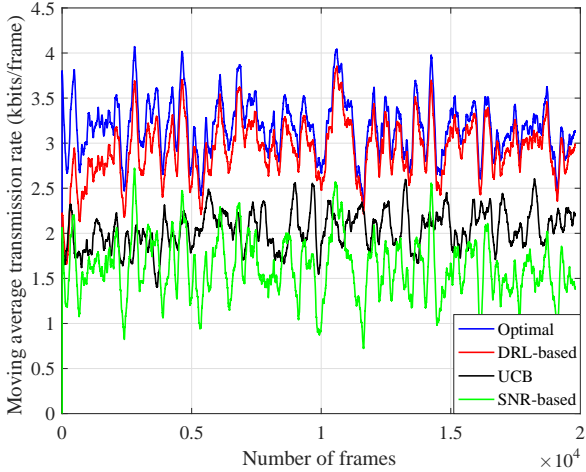}
\caption{Transmission rate comparison in a quasi-static interference scenario. Each value is a moving average of the previous $200$ frames and each curve is the average of $20$ trials.}
\label{Transmission_rate_quasi_static_interference}
\end{figure}

Fig. \ref{Transmission_rate_quasi_static_interference} compares the transmission rates of different algorithms in a quasi-static interference scenario. We consider two pairs of secondary users, i.e., namely, (ST-1, SR-1) and (ST-2, SR-2), although the proposed algorithm can handle more than two pairs of secondary users. The wireless links from the PU to both STs are completely blocked and STs cannot detect PU's signal at all. As such, we set the miss-detection probability of each ST to be $1$. Meanwhile, we assume that the correlation coefficient of the Rayleigh fading in two successive frames is $0.99$. In this scenario, the interference from each ST to the BS changes slowly. In the simulation, we set that the received average SNR $\frac{p_{\text{p}}\bar g_{\text{p}}}{\sigma^2}$ of the PU signal at the BS to $20$ dB and each received average interference-to-noise ratio $\frac{p_{\text{k}}\bar g_{\text{k}}}{\sigma^2}$ ($k\in \{1, 2\}$) at the BS to $5$ dB. From the figure, the optimal transmission rate fluctuates around $3$ kbits/frame, the transmission rate of the UCB learning algorithm increases from around $1.8$ kbits/frame to around $2.1$ kbits/frame, and the transmission rate of the SNR-based algorithm varies between $1$ kbits/frame and $1.4$ kbits/frame. Meanwhile, the proposed DRL-based MCS selection algorithm gradually achieves the optimal transmission rate of the optimal MCS selection algorithm, and is around $50\%$ higher than that of the UCB learning algorithm, and is $100\%$ higher than that of the SNR-based algorithm. This figure indicates that the proposed DRL-based algorithm is able to learn almost the perfect information of the quasi-static interference.

\begin{figure}[!t]
\centering
\includegraphics[scale=0.6]{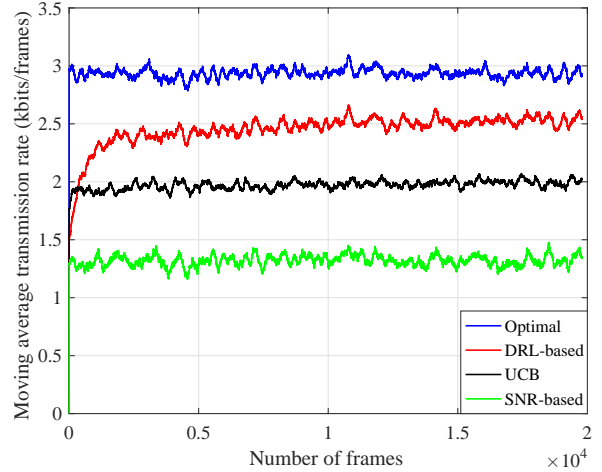}
\caption{Transmission rate comparison in a dynamic interference scenario. Each value is a moving average of the previous $200$ frames and each curve is the average of $20$ trials.}
\label{Transmission_rate_dynamic_interference}
\end{figure}

Fig. \ref{Transmission_rate_dynamic_interference} illustrates the transmission rates of different algorithms in a dynamic-interference scenario. We consider three secondary users, i.e., namely, (ST-1, SR-1), (ST-2, SR-2), and (ST-3, SR-3). The wireless links from the PU to ST-1 and ST-2 are completely blocked and ST-1/ST-2 cannot detect PU's signal at all, and the wireless link from the PU to ST-3 is not completely blocked but is extremely weak. As such, we set the miss-detection probabilities of the three STs to be $1$, $1$, and $0.5$, respectively. Meanwhile, we set the correlation coefficient of the Rayleigh fading in two successive frames to be $0$. In this scenario, the interference from each ST to the BS changes rapidly. In the simulation, we set that the received average SNR $\frac{p_{\text{p}}\bar g_{\text{p}}}{\sigma^2}$ of the PU signal at the BS is $20$ dB and each received average interference-to-noise ratio $\frac{p_{\text{k}}\bar g_{\text{k}}}{\sigma^2}$ ($k\in \{1, 2, 3\}$) at the BS is $5$ dB. From the figure, the optimal transmission rate is around $2.9$ kbits/frame, the transmission rate of the UCB learning algorithm converges to around $2$ kbits/frame, and the transmission rate of the SNR-based algorithm is around $1.3$ kbits/frame. Meanwhile, the transmission rate of the proposed DRL-based MCS selection algorithm converges to around $2.6$ kbits/frame, and is around $90\%$ of the optimal transmission rate, and is around $30\%$ higher than the transmission rate of the UCB learning algorithm, and is $100\%$ higher than that of the SNR-based algorithm. This figure verifies the effectiveness of the proposed DRL-based algorithm when the interference from STs to the BS is highly dynamic.

\subsection{Performance of the proposed algorithm with different $\Phi$}

\begin{figure}[!t]
\centering
\includegraphics[scale=0.6]{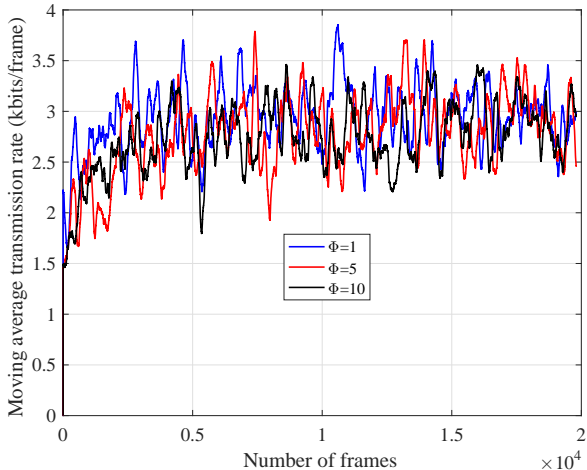}
\caption{Transmission rate of the proposed algorithm with different $\Phi$ in a quasi-static interference scenario. Each value is a moving average of the previous $200$ frames and each curve is the average of $20$ trials.}
\label{Quasi_static_memory}
\end{figure}

Fig. \ref{Quasi_static_memory} illustrates the transmission rate of the proposed algorithm with different $\Phi$ in a quasi-static interference scenario. In the simulation, the receive SNR of the PU signal at the BS, the number of secondary users, the miss-detection probability of each ST, and each receive interference-to-noise ratio at the BS are the same as those in Fig. \ref{Transmission_rate_quasi_static_interference}. Besides, we set $\Phi=1$, $\Phi=5$, and $\Phi=10$. From the figure, the transmission rate of the proposed algorithm remains almost the same when $\Phi$ increases from $1$ to $10$. The reason is as follows. The interference pattern from STs to the BS is dominated by the variation pattern of the corresponding channel gains. Since each interference channel gain changes slowly in a quasi-static interference scenario, the historical data in multiple previous frames provides almost the same interference pattern information as the historical data in the last frame for the DRL agent to infer the interference in the future. This figure indicates that it is unnecessary to put the historical data in multiple previous frames in each state when the interference from STs to the BS is quasi-static.

\begin{figure}[!t]
\centering
\includegraphics[scale=0.6]{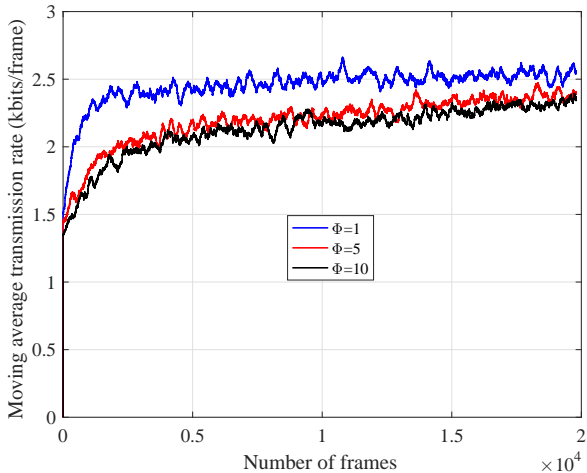}
\caption{Transmission rate of the proposed algorithm with different $\Phi$ in a dynamic interference scenario. Each value is a moving average of the previous $200$ frames and each curve is the average of $20$ trials.}
\label{Dynamic_memory}
\end{figure}

Fig. \ref{Dynamic_memory} investigates the transmission rate of the proposed algorithm with different $\Phi$ in a dynamic interference scenario. In the simulation, the receive SNR of the PU signal at the BS, the number of secondary users, the miss-detection probability of each ST, and each receive interference-to-noise ration at the BS are the same as those in Fig. \ref{Transmission_rate_dynamic_interference}. Besides, we set $\Phi=1$, $\Phi=5$, and $\Phi=10$. From the figure, the transmission rate of the proposed algorithm decreases as $\Phi$ increases from $1$ to $10$. The reason is as follows: As aforementioned, the interference pattern from STs to the BS is dominated by the variations of the corresponding channel gains. Since the channel model in the considered system is a first-order Markov process, each interference channel gain is only related to the interference channel gain in the previous frame. Note that each interference channel gain varies rapidly in a dynamic interference scenario. Then, the historical data in multiple previous frames cannot provide more interference pattern information than the historical data in the last frame for the DRL agent to infer the interference in the future, but in turn causes confusions to the DRL agent. As such, the DRL agent needs more frames to extract useful information about the interference pattern and infer the interference in the future. This figure indicates that it is harmful to put the historical data in multiple previous frames at each state when the interference from STs to the BS is highly dynamic.

\subsection{Balance between transmission rate and system overheads}

\begin{figure}[!t]
\centering
\includegraphics[scale=0.6]{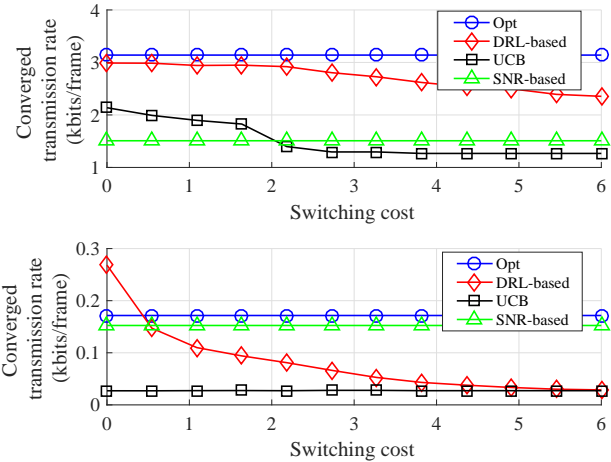}
\caption{Converged transmission rates and switching rates of different algorithms in a quasi-static interference scenario. }
\label{Switching_cost_quasi_static}
\end{figure}

Fig. \ref{Switching_cost_quasi_static} provides the converged transmission rates and the switching rates with different switching costs in a quasi-static interference scenario. In the simulation, the receive SNR of the PU signal at the BS, the number of secondary users, the miss-detection probability of each ST, and each receive interference-to-noise ration at the BS are the same as those in Fig. \ref{Transmission_rate_quasi_static_interference}. For a given switching cost, each algorithm runs $20,000$ frames similar to Fig. \ref{Transmission_rate_quasi_static_interference}. The converged transmission rate is obtained by averaging the latest $5000$ moving average transmission rates and the switching rate is $\frac{N_{\text{switching}}}{5000}$, where $N_{\text{switching}}$ is the number of switchings in the latest $5000$ frames. In this figure, the converged transmission rates and the switching rates of the optimal algorithm and the SNR-based algorithm remain constant as the switching cost $c$ grows. This is reasonable since the switching cost has no impact on both algorithms. Besides, The converged transmission rate of the UCB algorithm decreases from around $2.15$ kbits/frame to around $1.4$ kbits/frames as the switching cost increases from $c=0$ to $c=6$, and the corresponding switching rate remains around $0.03$. The converged transmission rate of the DRL algorithm decreases from around $3$ kbits/frame to around $2.4$ kbits/frames as the switching cost increases from $c=0$ to $c=6$, and the corresponding switching rate decreases from around $0.26$ to around $0.03$.

Fig. \ref{Switching_cost_quasi_static} indicates that, by adjusting the switching cost $c$, the DRL-based algorithm can achieve a higher converged transmission rate and a lower switching rate simultaneously than the SNR-based algorithm. For instance, when the switching cost is between $0.5$ and $6$, the converged transmission rate of the DRL-based algorithm is always higher than that of the SNR-based algorithm. Meanwhile the switching rate of the DRL-based algorithm is always lower than that of the SNR-based algorithm. Besides, by adjusting the switching cost $c$, the DRL-based algorithm can achieve a larger converged transmission rate than that of the UCB algorithm with a comparable switching rate. For instance, when the switching cost is between $4$ and $6$, the converged transmission rate of the DRL-based algorithm is always higher than that of the UCB algorithm, and the switching rates of both algorithms are almost identical. Therefore, when the interference from STs to the BS is quasi-static, the DRL-based algorithm can achieve a better balance between the primary transmission rate and system overheads than those of the optimal algorithm, the UCB algorithm, and the SNR-based algorithm.

\begin{figure}[!t]
\centering
\includegraphics[scale=0.6]{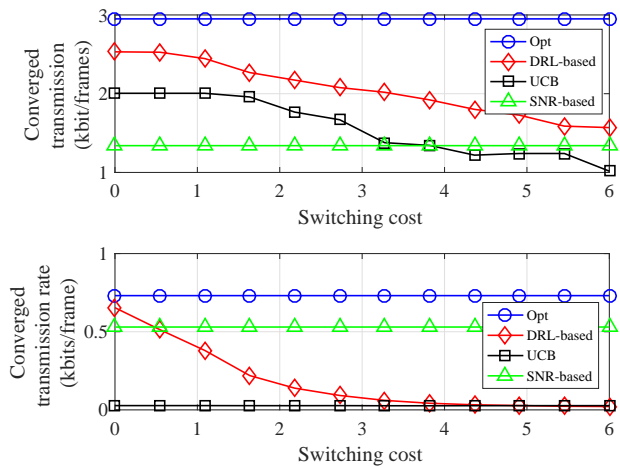}
\caption{Converged transmission rates and switching rates of different algorithms in a dynamic interference scenario.}
\label{Switching_cost_dynamic}
\end{figure}

Fig. \ref{Switching_cost_dynamic} provides the converged transmission rates and the switching rates with different switching costs in a dynamic interference scenario. In the simulation, the receive SNR of the PU signal at the BS, the number of secondary users, the miss-detection probability of each ST, and each receive interference-to-noise ratio at the BS are the same as those in Fig. \ref{Transmission_rate_dynamic_interference}. The converged transmission rate and switching rate are calculated with the same method as that in Fig. 8. In general, the trend of each curve in Fig. \ref{Switching_cost_dynamic} is similar to that in Fig. \ref{Switching_cost_quasi_static}. Specifically, by adjusting the switching cost $c$, the DRL-based algorithm can achieve a higher converged transmission rate and a lower switching rate simultaneously than those of the SNR-based algorithm. For instance, when the switching cost is between 0.5 and $6$, the converged transmission rate of the DRL-based algorithm is always higher than that of the SNR-based algorithm, and the switching rate of the DRL-based algorithm is always lower than that of the SNR-based algorithm. Besides, the converged transmission rate of the UCB algorithm ranges from $1.25$ kbit/frame to $2$ kbits/frame with a constant switching rate around $0.03$. The performance of the UCB algorithm can be achieved by the DRL-based algorithm through adjusting the switching cost $c$ between $c=3.5$ and $c=6$. Additionally, the DRL-based algorithm can also achieve a converged transmission rate higher than $2$ kbits/frame with a switching rate higher than $0.03$ when the switching cost $c$ is between $c=0$ and $c=3.5$. In other words, when the interference from STs to the BS is highly dynamic, the DRL-based algorithm can achieve a converged transmission rate similar to the UCB algorithm for a tight switching rate constraint scenario, and achieve a higher converged transmission rate than the UCB algorithm for a loose switching rate constraint scenario. To summarize, when the interference from STs to the BS is dynamic, the DRL-based algorithm can achieve a better balance between the primary transmission rate and system overheads than the optimal algorithm, the UCB algorithm, and the SNR-based algorithm.


\section{Conclusions}

In this paper, we studied a cognitive HetNet and proposed an intelligent DRL-based MCS selection algorithm for the PR to learn the interference pattern from STs. With the learnt interference pattern, the DRL agent at the PR can infer the interference in the future frames and select a proper MCS to enhance the primary transmission rate. Besides, we took the system overhead caused by MCS switchings into consideration and introduced a switching cost factor in the proposed algorithm to balance the primary transmission rate and system overheads. Simulation results showed that, the transmission rate of the proposed algorithm without the switching cost is $90\%\sim 100\%$ to that of the optimal MCS selection scheme and is $30\%\sim 100\%$ higher than those of benchmark algorithms. Meanwhile, the proposed algorithm with the switching cost can achieve better balances between the transmission rate and system overheads than both the optimal algorithm and benchmark algorithms.

\end{document}